\documentclass[10pt,journal]{./IEEEtran}
\usepackage{graphicx}
\usepackage{braket}
\usepackage[usenames,dvipsnames]{xcolor}
\usepackage{multirow}
\usepackage{booktabs}
\usepackage{tabularx}
\usepackage{makecell}
\usepackage{array}
\newcolumntype{Z}{ >{\centering\arraybackslash}X }
\newcolumntype{Y}{ >{\raggedleft\arraybackslash}X }
\newcolumntype{W}{ >{\raggedright\arraybackslash}X }

\makeatletter
\let\MYcaption\@makecaption
\makeatother
\usepackage[font=footnotesize]{subcaption}
\makeatletter
\let\@makecaption\MYcaption
\makeatother
\usepackage[utf8]{inputenc}

\usepackage{hyperref}

\begin{document}

\title{Rediscovery Datasets: Connecting Duplicate Reports }

\author{\IEEEauthorblockN{Mefta Sadat\IEEEauthorrefmark{1}, Ayse Basar Bener\IEEEauthorrefmark{2}, Andriy V. Miranskyy\IEEEauthorrefmark{3}}
\\
\IEEEauthorblockA{\IEEEauthorrefmark{1}\IEEEauthorrefmark{3} Dept. of Computer Science, Ryerson University, Toronto, Canada}
\\
\IEEEauthorblockA{\IEEEauthorrefmark{2}Dept. of Mechanical and Industrial Engineering, Ryerson University, Toronto, Canada}
\\
\IEEEauthorblockA{\{mefta.sadat, ayse.bener, avm\}@ryerson.ca}
}

\maketitle
\begin{abstract}
The same defect can be rediscovered by multiple clients, causing unplanned outages and leading to reduced customer satisfaction. In the case of popular open source software, high volume of defects is reported on a regular basis. A large number of these reports are actually duplicates / rediscoveries of each other. Researchers have analyzed the factors related to the content of duplicate defect reports in the past. However, some of the other potentially important factors, such as the inter-relationships among duplicate defect reports, are not readily available in defect tracking systems such as Bugzilla. This information  may speed up bug fixing, enable efficient triaging, improve customer profiles, etc.

In this paper, we present three defect rediscovery datasets mined from Bugzilla. The datasets capture data for  three groups of open source software projects: Apache, Eclipse, and KDE. The datasets contain information about approximately 914 thousands of defect reports over a period of 18 years (1999-2017) to capture the inter-relationships among duplicate defects. We believe that sharing these data with the community will help researchers and practitioners to better understand the nature of defect rediscovery and enhance the analysis of defect reports. 

\end{abstract}

\IEEEpeerreviewmaketitle

\section{Introduction}\label{sec:intro}
Software engineering research community mines bug repositories to conduct research in various areas. For example, one can detect duplicate reports to speed up report triaging (deduplication)~\cite{runeson2007detection, alipour2013contextual} and identification of the root cause of failure~\cite{bettenburg2008duplicate}, or to predict defect rediscoveries in order to proactively eliminate them before a customer finds~\cite{adams1984optimizing}, or to improve resource allocation to optimally manage the workforce~\cite{miranskyy2011metrics}, or to predict bug priority to improve planning~\cite{tian2013drone}, or to build customer profiles to improve quality assurance processes~\cite{miranskyy2009profiling}, or to automatically assign defect reports to owners to speed up time-to-fix of defects~\cite{anvik2006should}.

All of the above researchers leverage information about duplicate reports. There are already datasets that contain some information about duplicate reports (e.g.,~\cite{hooimeijer2007modeling, lamkanfi2013eclipse, 
herraiz2008towards, anvik2006should}). However, to the best of our knowledge, no recent datasets containing information on inter-relations between duplicate reports is available.  Thus, our \textbf{goal} is to create a collection of such datasets and share them with the community so that further research on duplicate defects can be performed. To achieve this goal, we mined bug repositories of  three groups of open source software projects (Apache, Eclipse, and KDE),
gathering information about duplicate defects, making it easy to identify  relations between all of the duplicate defect reports. The datasets contains information about $\approx 914$ thousands defects that have been reported in the last 15-18 years (depending on the project). The resulting datasets are located at \url{http://doi.org/10.5281/zenodo.400614} \cite{sadat_mefta_2017_400614}.  

Throughout the paper the following \textbf{terminology} (adopted from \cite{bettenburg2008duplicate, runeson2007detection, miranskyy2009profiling}) is used. Original defect \textit{discovery} can be defined as the moment when a customer encounters a defect in the software for the very first time. Encounter is manifested by a problem or a fault in the software that leads to an undesired outcome or even a software \textit{failure}. The customer then submits a \textit{report} to a bug tracking system describing the problem.

\begin{figure}[t]
    \centering
    \includegraphics[width=\columnwidth]{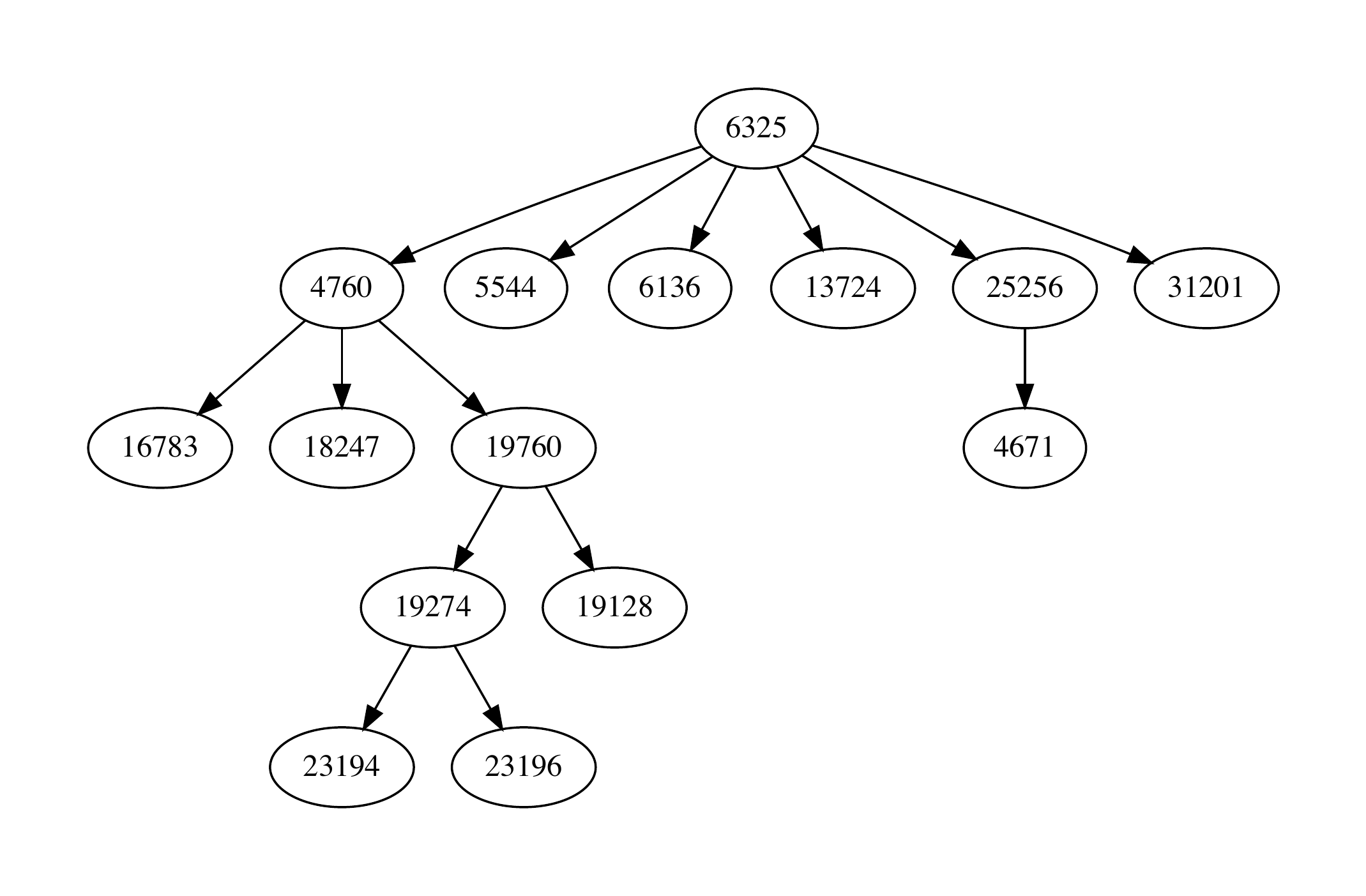}
    \caption{Graph  of rediscoveries of Eclipse report $\#4671$. Report $B$ being duplicate of report $A$ is denoted by $A \rightarrow B$.  Note that even though report $\#4671$ is the original discovery, a later report $\#6325$ was chosen by developers as the master report. We can say that the failure associated with report $\#4671$ was discovered $15$ times in total (counted as the total number of vertices/reports in the graph) and rediscovered $14$ times (total number of duplicate reports). }
    \label{fig:dup_graph}
\end{figure}

If another\footnote{In general, this can be the same customer, if they have multiple copies of the product.}
customer encounters the same defect again, it is called defect \textit{rediscovery}. This customer will then submit a new report to the bug tracking system.  During report triaging, developers identify if a new report relates to a discovery of a new defect or to a rediscovery of an existing one. If it is a rediscovery, then developers typically mark the most recent report as a duplicate and link it to the original report (in some cases the link may be established incorrectly: ``to err is human'').  They then choose one of the linked reports as a \textit{master report} and the rest of the reports associated with this particular failure will be deemed \textit{duplicates} of the master report. Note that the report associated with the first discovery does not necessarily become a master report -- sometimes developers choose a report of one of the rediscoveries as a master one. Given that there can be more than one rediscovery of the same defect, the network linking the original report with duplicate ones (which we call the \textit{graph of rediscoveries}) may become complex. For example, Figure~\ref{fig:dup_graph} shows the graph of rediscoveries for Eclipse report $\#4671$. Note that the master report in this case is not the original report. 

Summing up original discovery and rediscovery count yields\textit{ total number or reports for a given failure}. If a given report was discovered in total once, then it means that it was never rediscovered; discovered twice -- means that it was rediscovered once, and so on. In the case of Figure~\ref{fig:dup_graph}, report $\#4671$ was rediscovered $14$ times. Thus, the total number of reports for a failure associated with report $\#4671$ is $15$.

The rest of the paper is organised as follows. Description of the datasets and our methodology for data extraction and transformation are given in Section~\ref{sec:data}, key features of the datasets -- in Section~\ref{sec:data_analysis}, possible uses and relevance to the research community -- in Section~\ref{sec:rel}, challenges and limitations -- in Section~\ref{sec:chal}, and conclusions -- in  Section~\ref{sec:sum}.

\begin{table}[]
\centering
\caption{Extracted Attributes}\label{table:attr}
\label{my-label}
\begin{tabular}{p{1.2cm}p{6.7cm}}
\toprule
\textbf{Attribute}    & \textbf{Definition}                                                                                                                                                                                                     \\
\midrule
id           & The unique integer identifying a report.                                                                                                                                                        \\
product      & The name of the software subsystem the report belongs to.                                                                                                                                                      \\
component    & The name of the component the report is associated with.                                                                                                                                                       \\
reporter     & The unique username of the person who opened the report.                                                                                                                                                       \\
bug\_status  & The current status of the report.                                                                                                                                                                              \\
resolution   & The current resolution of the report.                                                                                                                                                                          \\
priority     & Represents how quickly the defect should be fixed.                                                                                                                                  \\
bug\_severity &  Defect's degree of impact on the whole system.                                                                                                                                 \\
version      & The version the defect was observed in.                                                                                                                                                                        \\
short\_desc  & A short textual summary of the report.                                                                                                                                                                         \\
opendate     & The date when the report was opened.                                                                                                                                                                           \\
dup\_list    & The list of ids of duplicates of a given report; if the report does not have any duplicates -- the value is an empty string.                                                                        \\
root\_id     & A derived attribute -- the id of the root vertex of the graph of rediscoveries, which typically resembles the master report. If the report does not have any duplicates -- the value is an empty string. \\
disc\_id     & A derived attribute -- the id of the oldest defect report (i.e., the one that is opened first) in the graph of rediscoveries. If the defect does not have any duplicates -- the value is an empty string.      \\
\bottomrule
\end{tabular}
\end{table}

\section{Methodology: extraction and transformation}\label{sec:data}

For each group of the software projects, the set of attributes that we extracted from each report are given in Table~\ref{table:attr}. We performed the following four extraction and transformation steps to obtain the attributes.

    \textbf{Step 1: Retrieval of report \textit{id}s.} 
    For each of the software projects we selected, we mined its Bugzilla defect tracking system which numbers defect reports sequentially  with an integer \textit{id}, with the first \textit{id} set to $1$.
    
    Given the sequential nature of the data, we query a given Bugzilla engine for reports opened within the last seven days (at the day of data gathering) and select the maximum \textit{id} value, denoted by $I_{\max}$ returned by the engine. Thus, for a given engine the range of reports \textit{id}s is set to $[ 1, I_{\max}]$.

    \textbf{Step 2: Data mining and extraction.} 
    The data were extracted using a custom-built web scraper. The input to the scraper was the range of \textit{id}s to be mined - identified in the previous step. The scraper outputs all the attributes mentioned in Table~\ref{table:attr} (except the two derived attributes) in \textit{CSV} format (one line per report), saving intermediate results, as the extraction process takes several days to complete. 
    
     \textbf{Step 3: Construction of the dataset.} 
     First, we aggregate all intermediate results for a given project in a single \textit{CSV} file. 
     
     Second, we eliminate rows from the \textit{CSV} file for which a report either does not exist or is not available. The former may happen because the report may get cancelled by a user before submission or may be erased by a bug tracker administrator. The latter may happen because we do not have sufficient permissions to access a given report. The former case cannot bias our dataset, as the data does not exist. However, the latter case may lead to bias, if the number of reports that we cannot access is large. We built a script that computed the number of \textit{id}s associated with each case (by analysing error messages returned by the bug tracking engine). Details of our analysis are provided  in Table~\ref{table:attr}.

     \textbf{Step 4: Construction of derived attributes.}
     In order to construct derived attributes, we built a directed graph $G$ linking \textit{id} with its duplicates using information stored in the \textit{dup\_list} attribute. Going back to example given in Figure~\ref{fig:dup_graph}, report $\#19274$ has two duplicates linked to it ($\#23194$ and $\#23196$), as per the \textit{dup\_list} attribute. Thus, we will add to the $G$ two edges:  $19274 \rightarrow 23194$ and $19274 \rightarrow 23196$. We repeat this process for each report in a given dataset. We then use Graphviz software~\cite{gansner00anopen} to identify all  `connected components' (in the graph theory sense of the term) in the $G$. The resulting connected components represent the graph of rediscoveries for each of the original defects. An example of such connected component is given in Figure~\ref{fig:dup_graph}. 
         
     We then analyze each graph of rediscoveries (connected component) and identify the root vertex  (typically, this report is a master report) and the vertex associated with the \textit{id} with the oldest \textit{opendate}. The former becomes \textit{root\_id} value for each report associated with a given graph of rediscoveries; the latter value becomes $disc\_id$. For example, in case of Figure~\ref{fig:dup_graph}, the \textit{root\_id} value for all the reports will be set to $6325$ and \textit{disc\_id} to $4671$ (since, by design of the Bugzilla defect tracking system, the smaller the defect \textit{id} -- the older the defect). 
    Then, we merge the original dataset with the derived attributes and store the resulting dataset in the \textit{CSV}, \textit{SQL}, and \textit{Neo4j} formats.

\section{Analysis of the dataset}\label{sec:data_analysis}

The summary statistics of the datasets are given in Table~\ref{tbl:summary_stats}. The number of reports that we gathered (column `Total accessible reports count') ranges from $\approx 44$ thousands for Apache to $\approx 504$ thousands for Eclipse. The reports were opened between years 1999 and 2017.

\begin{table*}[!ht]
\setlength\tabcolsep{4pt}
\caption{Summary statistics.}\label{tbl:summary_stats}
\begin{tabularx}{\textwidth}{ZYYYYZZYYYY}
\toprule
  Project name   &  
 \multicolumn{1}{Z}{ Total accessible reports count } &  
 \multicolumn{1}{Z}{ Inaccessible reports count } &  
 \multicolumn{1}{Z}{ Rediscoveries count } & 
 \multicolumn{1}{Z}{ Distinct \textit{disc\_id} count } &  
 Min report \textit{opendate} \tiny{(YYYY-MM-DD)}&  
 Max report \textit{opendate} \tiny{(YYYY-MM-DD)} & 
 \multicolumn{1}{Z}{ Max number of rediscoveries } &
 \multicolumn{1}{Z}{ Distinct \textit{product}s count } & 
 \multicolumn{1}{Z}{ Distinct \textit{product-component}s count} & 
 \multicolumn{1}{Z}{ Non-rediscovered reports (\% of total) } \\ 
 \midrule

Apache  & 44,049  & 1 & 3,616 & 2,416 & 2000-08-26 &  2017-02-10 & 19 & 35 & 350 & 86 \\ 
Eclipse  & 503,935  & 560 & 52,499 & 31,811 & 2001-10-10 &  2017-02-07 & 128 & 232 & 1,486 & 83 \\ 
KDE  & 365,893 &  4,818 & 82,359 & 26,114 & 1999-01-21 &  2017-02-13 & 405 & 584 & 2,054 & 70 \\

\bottomrule
\end{tabularx}
\end{table*}

As discussed in Section~\ref{sec:data}, we could not access some of the reports. The percentage of such reports (shown in column `Inaccessible reports count') is small: $0.002\%$  for Apache, $0.1\%$  for Eclipse, and $1.3\%$ for KDE. These reports also lead to $1$, $79$, and  $33$ inaccessible edges in $G$ for Apache, Eclipse, and KDE, respectively. Thus, these missing observations should not bias the datasets significantly and can be ignored. 

To gather information about original discoveries and rediscoveries of reports, as discussed in Section~\ref{sec:data}, we analysed graphs of rediscoveries (similar to the one shown in Figure~\ref{fig:dup_graph}). Such graphs can become fairly large: based on Table~\ref{tbl:summary_stats}, the maximum number of rediscoveries of an original report (per graph of rediscoveries) ranges  from $19$ for Apache to $405$ for KDE. Most graphs are acyclical, with the exception of one cyclical graph in Eclipse and four in KDE datasets.

\begin{figure}
    \centering
    \begin{subfigure}{\columnwidth}
    \includegraphics[width=1.0\linewidth]{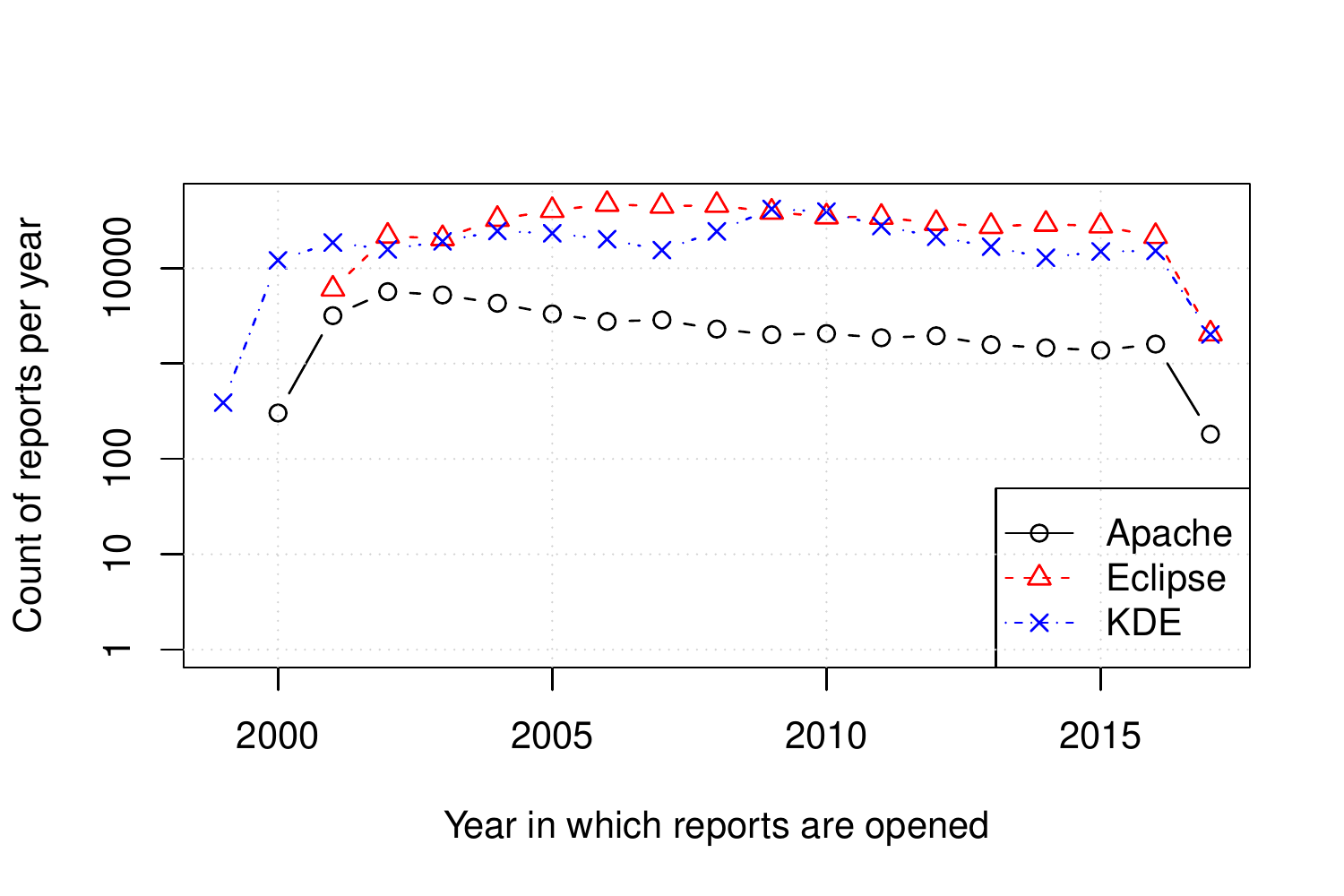}
    \caption{Number of reports per year.}
    \label{fig:per_year_count_all}
    \end{subfigure}
    \begin{subfigure}{\columnwidth}
    \includegraphics[width=1.0\linewidth]{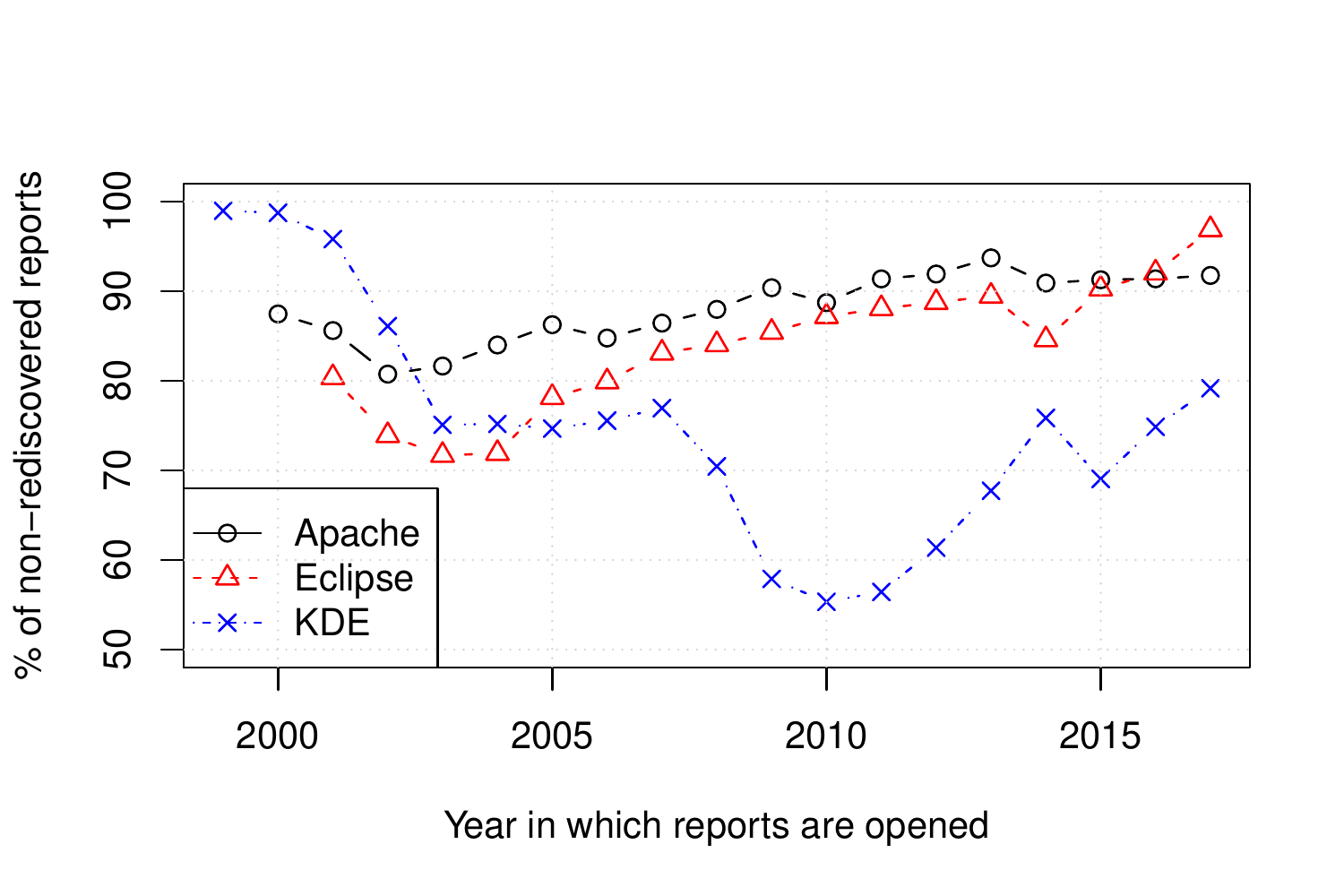}
    \caption{Percent of reports that have not been (yet) rediscovered.}
    \label{fig:per_year_count_nonredisc}
    \end{subfigure}
    \caption{Per-year analysis. The data are current as of February 2017, thus the dataset for year 2017 is incomplete, hence the ``dip'' in reports for year 2017. By construction, zero observations for a given year are not shown.}
    \label{fig:per_year}
\end{figure}

\begin{figure}[ht]
    \centering
    \includegraphics[width=\columnwidth]{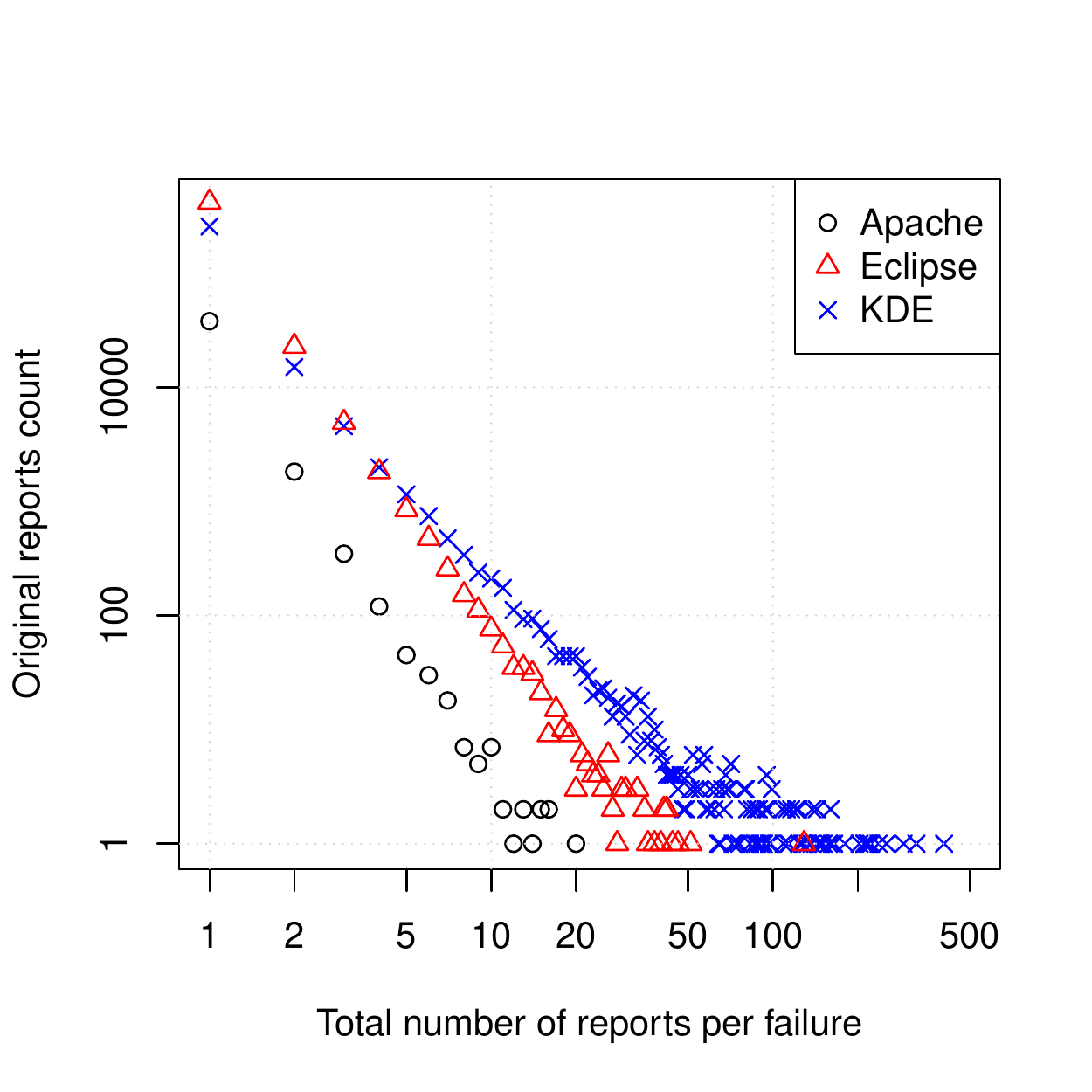}
    \caption{Count of the total number or reports for a given failure vs. count of original reports. If a given failure was reported once, then it means that it was never rediscovered; reported twice -- means that it was rediscovered once, and so on (see Section~\ref{sec:intro} for details). For example, Apache dataset has $38017$ reports that were never rediscovered (i.e., discovered once) and $1825$ reports that were rediscovered once (i.e., discovered twice).}
    \label{fig:disc_distr}
\end{figure}

The percentage of the original reports that were rediscovered at least once  ranges from $5\%$ ($2416 / 44049$) for Apache to $7\%$ ($26114 / 365893$) for KDE. The distributions of the total number of reports (obtained by combining rediscovery and original defect count, as discussed in Section~\ref{sec:intro}) for a given failure are given in Figure~\ref{fig:disc_distr}. The distributions are heavy-tailed as evident from the linear structure of the data plotted on the log-log scale.
The number of reports per year changes, as seen in Figure~\ref{fig:per_year_count_all}. Magnitude-wise, the number of reports ranges from thousands for Apache to tens of thousands for Eclipse and KDE (with the exception of the first and last reporting year for each project).

Overall, percentage of reports that are not rediscovered ranges between $70\%$ for KDE and $86\%$ for Apache. However, these values change from year to year, as shown in Figure~\ref{fig:per_year_count_nonredisc}. This figure may suggest that for the last seven years percentage of non-rediscovered reports grows up (albeit non-monotonically). For example, for defects opened in 2016, the percentage of non-rediscovered defects ranges from $75\%$ for KDE to $92\%$ for Eclipse (compare these numbers with the average values of $70\%$ and $86\%$, respectively). 

\begin{figure}
    \centering
    \includegraphics[width=\columnwidth]{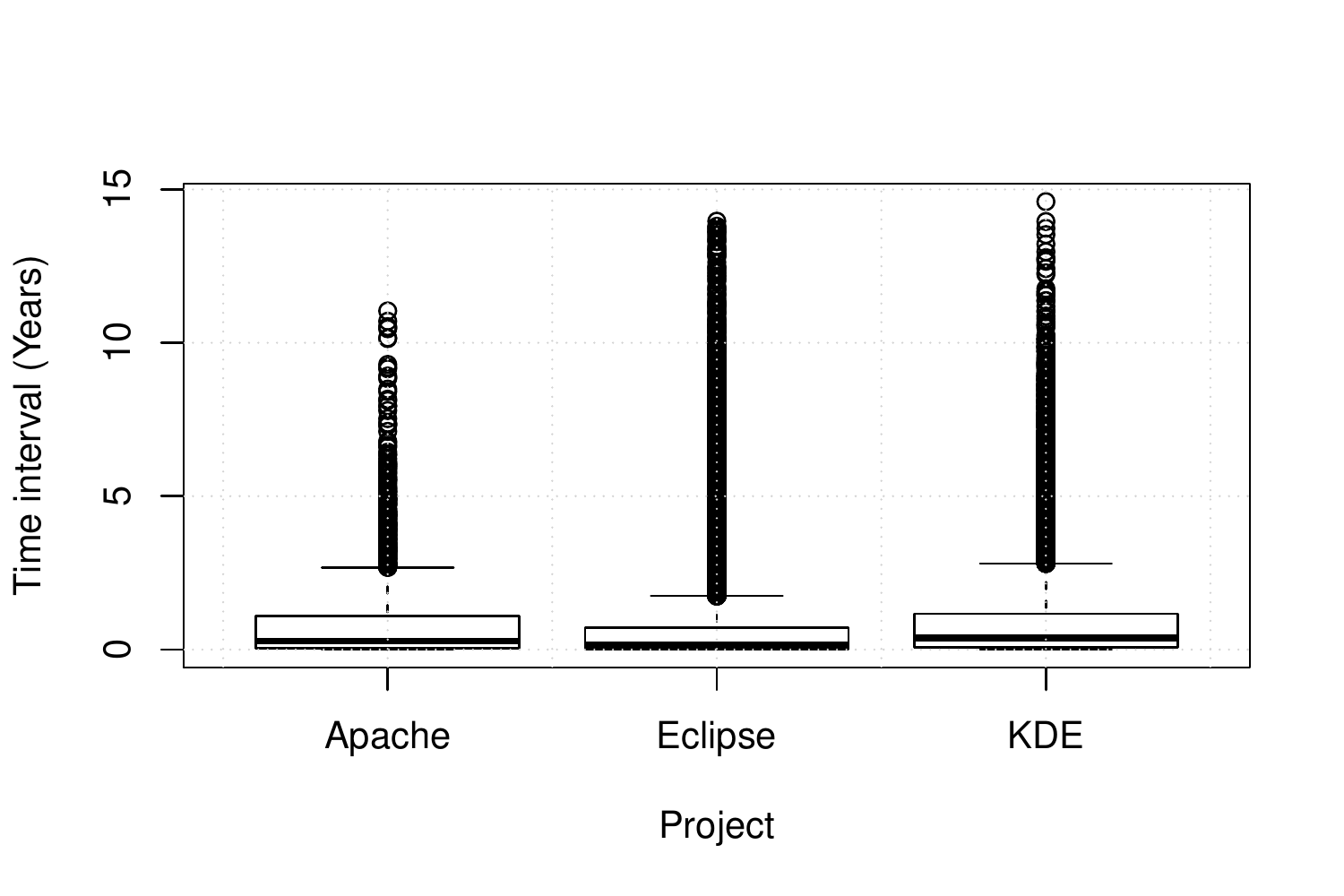}
    \caption{Distributions of time intervals between the original discovery and the latest rediscovery for a given graph of rediscoveries.   }
    \label{fig:redisc_time_int}
\end{figure}

However, in the future, users may encounter and report some of the defects discussed in these non-rediscovered reports. This will lead to reduction of the number of non-rediscovered reports opened in previous years. To confirm this conjecture, we plot the distribution of time intervals between the opening dates of the original discovery and the latest rediscovery, shown in Figure~\ref{fig:redisc_time_int}. The figure suggests that some reports get rediscovered years after the original discovery. Even for the graph of rediscoveries shown in Figure~\ref{fig:dup_graph}, the time interval between open dates of the original report $\#4671$ and its latest rediscovery $\#31201$ was $\approx 1.3$ years.

\begin{figure}
    \centering
    \includegraphics[width=\columnwidth]{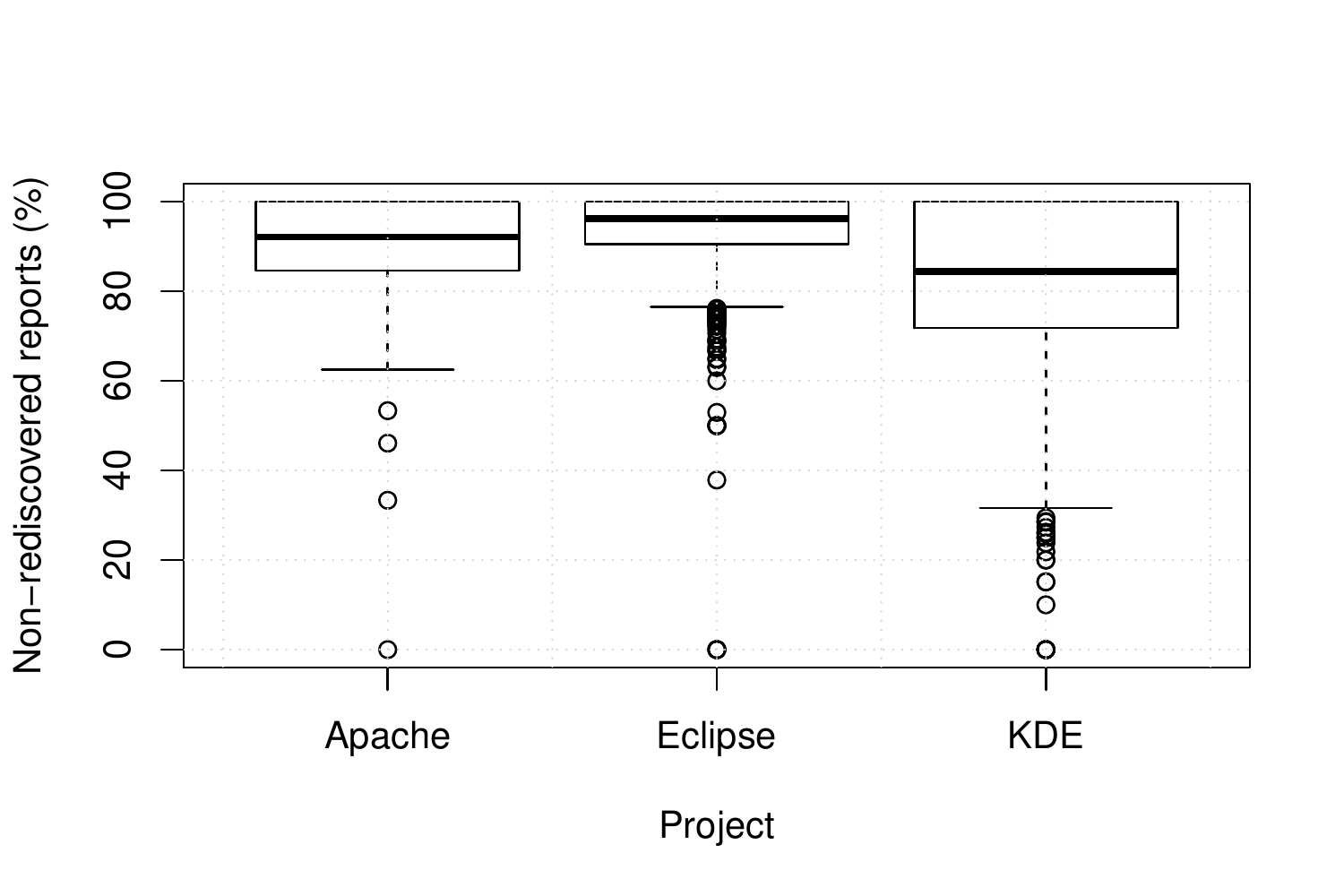}
    \caption{Distribution of non-rediscovered reports per \textit{product-component}.}
    \label{fig:per_comp_never_redisc}
\end{figure}

The number of \textit{product}s per project ranges from $35$ for Apache to $584$ for KDE; the number of \textit{product-component} tuples per project -- from $350$ for Apache to $2054$ for KDE. The percentage of reports that are not rediscovered per product-component is given in  Figure~\ref{fig:per_comp_never_redisc}. The median percentage ranges between $84\%$ for KDE to $96\%$ for Eclipse. However, there are outliers with low percentage of non-rediscovered defects, suggesting that different components may exhibit different behaviour. Therefore, various \textit{product-component}s may be studied independently.

\section{Relevance of the Dataset}\label{sec:rel}
Based on the analysis of the datasets given in Section~\ref{sec:data_analysis}, we believe that these datasets provide a rich ground for researchers interested in analyzing defects for various purposes discussed in Section~\ref{sec:intro}. For example, they can be used for cross-product verification of the models built by researchers to speed up triaging and identification of root causes of failures,  predict defect rediscoveries, or assign owners to reports. The data are provided in the \textit{CSV}, \textit{SQL}, and \textit{Neo4j} formats, enabling easy investigation of the datasets.

\section{Challenges and Limitations}\label{sec:chal}
We do not have access to a number of reports, which may bias our dataset (as discussed in Section~\ref{sec:data_analysis}). However, given that the percentage of such reports is low ($0.002\%$  for Apache, $0.1\%$ for Eclipse, and $1.3\%$  for KDE), the  dataset should not be affected significantly.

Our list of attributes does not cover all of the defect reports' attributes available in Bugzilla. However, our dataset helps researchers to narrow down a set of the defect reports that have to be mined to gather such additional attributes (e.g., comments associated with a given defect report). For example, if  researchers are interested in the analysis of duplicate defects of Eclipse dataset, they can focus on mining just $17\%$ $((52499 + 31811)/ 503935)$ of the reports (as shown  in Table~\ref{tbl:summary_stats}), with report \textit{id}s being readily available in our datasets. Thus, this would allow them to save time and computational resources on the costly extraction and transformation process.

In addition, some of the reports that are currently non-rediscovered may be rediscovered in the future  (as discussed in Section~\ref{sec:data_analysis}). This has to be taken into consideration during data analysis.

\section{Summary}\label{sec:sum}
In this paper, we present datasets collected from three groups of projects (Apache, Eclipse, and KDE), aimed at capturing information associated with duplicate / rediscovered defects. We describe the schema of the datasets, extraction and transformation process, and present analysis of the datasets. We believe that these datasets will aid researchers and practitioners in gathering insight into usage of duplicate reports in various areas of software engineering.

\section*{Acknowledgment}
\small{This work is partially supported by NSERC grants 402003-2012 and RGPIN-2015-06075.}

\bibliographystyle{abbrv}
\bibliography{ref}

\end{document}